\documentclass[useAMS,usenatbib,usegraphicx]{mn2e}

\title[New Constraints on 298 Baptistina]{New Constraints on the Asteroid 298 Baptistina, the Alleged Family Member of the K/T Impactor}
\author[]
{D.~J. Majaess$^{1,4}$, D. Higgins$^{2}$, L.~A. Molnar$^{3}$, M.~J. Haegert$^{3}$, D.~J. Lane$^{1,4}$,
\newauthor  D.~G. Turner$^{1}$ and I. Nielsen$^{5}$ \\ \\
  $^1$Department of Astronomy and Physics, Saint Mary's University, Halifax, Nova Scotia, Canada (dmajaess@ap.smu.ca) \\
  $^2$The Hunters Hill Observatory, Canberra, Australia (higginsdj@bigpond.com) \\
  $^3$Calvin College, Grand Rapids, Michigan, USA (lmolnar@calvin.edu) \\
  $^4$The Abbey Ridge Observatory, Stillwater Lake, Nova Scotia, Canada (dlane@ap.smu.ca) \\
  $^5$Hamburger Sternwarte Observatory, Hamburg, Germany (inga.nielsen@gmx.de) \\
}
\date{Accepted for Publication in the Journal of the \textbf{Royal Astronomical Society of Canada} (\textit{JRASC}) \\ Accepted 2008 November $1^{st}$; Received 2008 June $20^{th}$}

\begin{document}

\maketitle

\begin{abstract}
In their study \citet{bo07} suggest that a member of the Baptistina asteroid family was the probable source of the K/T impactor which ended the reign of the Dinosaurs 65 Myr ago.  Knowledge of the physical and material properties pertaining to the Baptistina asteroid family are, however, not well constrained.  In an effort to begin addressing the situation, data from an international collaboration of observatories were synthesized to determine the rotational period of the family's largest member, asteroid 298 Baptistina ($P_R=16.23\pm0.02$ hrs).  Discussed here are aspects of the terrestrial impact delivery system, implications arising from the new constraints, and prospects for future work. 
\\ \\
Dans leur recherche, \citet{bo07} sugg{\`e}re qu'un ast{\'e}ro{\"i}de membre de la famille des Baptistina, est la source probable de l'impacte du K/T qui a mis fin au r{\`e}gne des dinosaures environ 65 milliards d'ann{\'e}es. La connaissance des propri{\'e}t{\'e}s physiques et mat{\'e}rielles de la famille d'ast{\'e}ro{\"i}de Baptistina n'est cependant pas bien {\'e}tudi{\'e}s. Dans un effort de comprendre la situation, des donn{\'e}es prises d'une collaboration internationale d'observatoires ont {\'e}t{\'e} synth{\'e}tis{\'e}es pour d{\'e}terminer la p{\'e}riode de rotation du plus grand membre de la famille, l'ast{\'e}ro{\"i}de 298 Baptstina ($P_R=16.23\pm0.02$ hrs). Cette {\'e}tude discute les aspects du syst{\`e}me de l'impact terrestriel, les implications soulev{\'e}es de ces nouvelles contraintes, et les opportunit{\'e}s de futures recherches.
\end{abstract}

\section*{Introduction}
Terrestrial impactors (asteroids and comets) have been suggested to play a major role in modulating the existence of life on Earth, as the dating of craters linked to kilometer-sized impactors at Popigai and Chesapeake Bay, Chicxulub \citep{hi93}, and Morokweng and Mjolnir strongly correlate in age with three of the last major global extinctions (late-Eocene, Cretaceous-Tertiary, and Jurassic-Cretaceous respectively).   Indeed, readers can view images of the corresponding impact craters at the \textit{Earth Impact Database} ( http://www.unb.ca/passc/ImpactDatabase/ ), which is maintained by the Planetary and Space Science Center at the University of New Brunswick. 

One of the challenges, undoubtedly, is to explain how such impactors transition from otherwise benign orbits in the solar system to become near-Earth objects (NEOs). Historically, it has been suggested that the cause of such extinctions may be linked to an influx of comets by means of a perturbation of the Oort Cloud, a spherical zone of loosely-bound comets thought to encompass the periphery of the solar system. A litany of possible causes have been put forth as catalysts for such a perturbation, most notably density gradients (stars and the interstellar medium) encountered as the Sun oscillates vertically through the plane of the Milky Way during its revolution about the Galaxy, or interactions with a suspected substellar companion to the Sun (Nemesis). Certain ideas are advocated because they inherently assume a periodicity to mass extinction events, although unproven, rather than stochastic punctuations. A different impact delivery system revisited below is based primarily on orbital resonances, and favours a reservoir of projectiles from the asteroid belt located between Mars and Jupiter, in addition to comets from the Kuiper Belt and Oort Cloud, the former extending beyond Neptune from 35 A.U. to 50+ A.U. 

\begin{figure}
\begin{center}
\includegraphics[width=8cm]{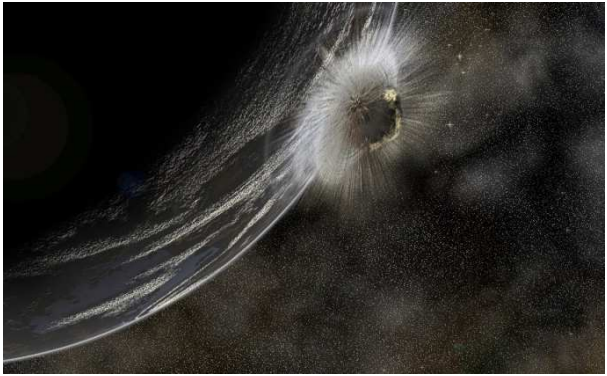}
\end{center}
\caption{A cataclysmic encounter between Earth and a large asteroid as envisioned by artist / astronomer Inga Nielsen.}
\label{fig1}
\end{figure}

\subsection*{Resonances}
Large bodies can be delivered from both Belts into Earth-crossing orbits by means of resonances (secular and mean motion).  Formally, a resonance occurs where the orbital periods of two bodies are commensurate (ratios of integers). For example, an asteroid that is near a 2:1 resonance with Mars will orbit the Sun once for each two orbits Mars completes. Most importantly, asteroids near resonances may experience periodic perturbations from a planet that could lead to an increasing eccentricity and a subsequent close encounter, resulting in the asteroid being gravitationally scattered. Observations confirm that areas in the Main Belt associated with strong resonances with the orbit of Jupiter (or Mars) are indeed devoid of asteroids (Kirkwood Gaps), securing the resonance phenomenon as a feasible mechanism for transporting objects from the Belt. Inevitably, a fraction of the asteroids depleted by orbital resonances become NEOs.

A distribution analogous to the Kirkwood Gaps is also noted in computational models of the Kuiper Belt \citep{ma04}, where Neptune plays a major role in scattering comets. Moreover, simulations confirm that comets from the region could then enter other planet-crossing orbits, although the relevant impact probabilities are difficult to constrain because firm statistics on the Kuiper Belt's comet population lie beyond present limits of solid observational data. The James Webb Space Telescope (JWST), scheduled for launch in $\simeq2014$, and the ongoing Canada France Ecliptic Plane Survey (CFEPS) should place firmer constraints on the Kuiper Belt demographic. Indeed, many Canadian astronomers and institutions are active partners in JWST (i.e. John Hutchings, NRC-HIA, Ren\'{e} Doyon, Universit\'{e} de Montr\'{e}al) and CFEPS (i.e. JJ Kavelaars and Lynne Jones, NRC-HIA, Brett Gladman, UBC).

\subsection*{The Yarkovsky Effect}
The Yarkovsky Effect (YE) is another component of the delivery system that can work to enhance the transport of asteroids (or comets) into resonances, essentially increasing the possibility that bodies not near resonances may eventually arrive at such locations. In its simplest form, the canonical YE arises from a temperature differential between the sunlit and dark sides of an object exposed to the Sun. Thermal energy from the object is therefore reradiated asymmetrically, causing the body to experience a thrust that may result in an outwards or inwards orbital migration, depending upon its sense of rotation and the direction of the resulting rocket force (see \citet{ru98} for details).

The YE also allows constraints to be placed on the ages of asteroid families \citep{vo06}---which is used below to connect the Bapstina asteroid family to the K/T impactor---although such a framework is still in its scientific infancy. The force causes smaller asteroids to undergo a greater orbital migration in comparison with larger bodies, producing a characteristic distribution in semi-major axis space \citep[see figure 1,][]{bo07}. Computer simulations can then determine at what time after the fragmentation of a parent body the present day distribution of an asteroid family is reproduced satisfactorily. Such analyses depend on knowledge of several different parameters, which in the case of the Baptistina asteroid family (BAF) are not well established as discussed below. Lastly, it is noted that the YE has been invoked to describe the motion of asteroid 6489 Golevka \citep{ch03}, and additional efforts to test and confirm the effect observationally are forthcoming.

\subsection*{Summary}
In sum, a three-component terrestrial impact delivery system could begin in the Belt with the fragmentation of a parent body near a resonance by means of a collision that spawns hundreds of smaller asteroids, thereby augmenting the statisticaly probability and likelihood of a terrestrial impactor. After fragmentation, a particular asteroid could then enter a nearby resonance or drift there by means of the YE, where it may be scattered by a planet into an Earth-crossing orbit. 

\begin{figure*}
\begin{center}
\includegraphics[width=12cm]{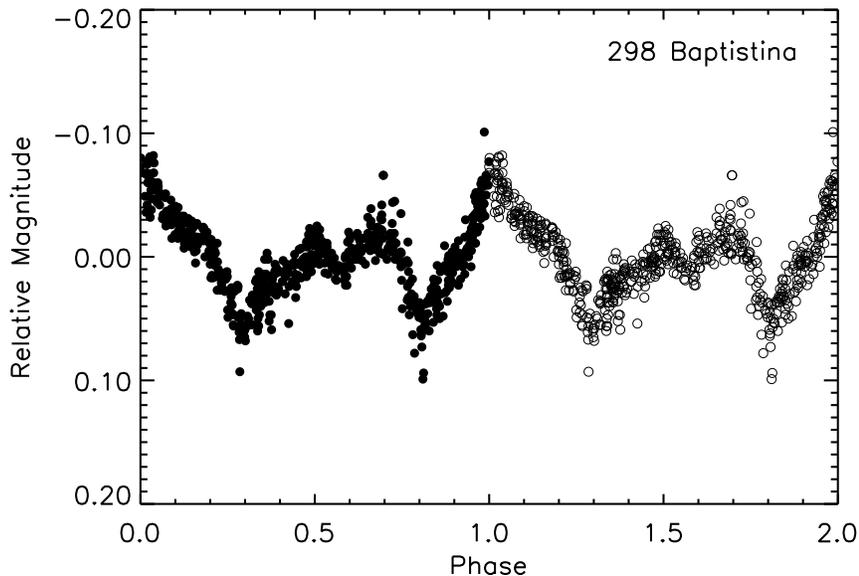}
\end{center}
\caption{The light curve for 298 Baptistina phased with a period of $P_R=16.23\pm0.02$ hours.}
\label{fig2}
\end{figure*}

\section*{The Alleged Baptistina/KT Impactor Connection}
In harmony with the delivery model revisited above, \citet{bo07} postulate that the Baptistina asteroid family formed from the catastrophic breakup of its progenitor approximately 160 Myr ago, following which some debris entered a nearby resonance and eventually led to the ejection of what would inevitably become the K/T impactor. Their proposal is argued, based in part (the reader is also referred to the comprehensive supplemental texts that accompany the \citet{bo07} paper), on the following lines of evidence: (i) the asteroid family is located near a resonance capable of delivering passing asteroids into planet-crossing orbits. (ii) The purported destruction of the parent body 160 Myr ago, an age inferred from sorting the asteroids into orbital parameter space according to the YE, created a prodigious supply of BAF members that inevitably populated the NEO demographic, consistent with an alleged increase in the terrestrial impact rate during the same era. It should be noted, however, that the terrestrial record of impacts suffers from poor statistics owing to the subsequent erosion of craters with time, complicating any statistical interpretation. (iii) The K/T impactor and BAF share a similar composition.  Yet the results of \citet{re08} appear to imply otherwise (at least vis {\`a} vis the family's largest member, 298 Baptistina), and moreover, the suggested C-type composition of BAF members would not be unique; such asteroids are found throughout the Belt.

The final conclusions are based on statistical grounds, namely what is the probability that the impactor was a fragment from the creation of the BAF rather than a random C-type asteroid (or background population). Bottke et al. (2007) suggest that there is a 90\% chance that the K/T impactor was a BAF member, or $1\pm1$ BAF members of size d$\ge10$ km impacted Earth in the past 160 Myr. Readers should be aware that contributions from the background population are difficult to assess. In addition, accurate modeling of the YE requires knowledge of both physical and material properties that are conducive to BAF members, sensitive parameters that are poorly constrained and require further research by the community at large. Indeed, the rotational period derived here for BAF member 298 Baptistina ($P_R=16.23\pm0.02$ hrs, see below) is a factor of three greater than the value used in the simulations, although \citet{bo07} adopted a value that may be consistent with smaller-sized  BAF members \citep[$P_R\simeq6\pm2$ hours,][]{ph00,pr02}. The difference in rotational periods noted above is sufficient to warrant additional investigations to confirm the mean rotational period and material properties conducive to kilometer-sized BAF members. Such work needs to be pursued in conjunction with increasing the number of known family members and reaffirming the family's taxonomy. Efforts to secure such parameters will invariably lead to stronger constraints on the properties of family members and might permit a more confident evaluation of whether the source of the K/T impactor was indeed a $\simeq10$ km sized BAF member. Lastly, and \textit{most importantly}, irrespective of the conclusion regarding the putative source of the K/T impactor, the approach outlined by \citet{bo07} provides the quantitative framework and a pertinent example needed to effectively characterize the terrestrial impact delivery system.

\section*{Observations}
Asteroid 298 Baptistina was discovered over a century ago, in September 1890, by the French astronomer Auguste Charlois. The origin of the asteroid's designation (Baptistina) is unknown, an uncertainty that is also representative of the asteroid's rotational period, morphology, size, etc. A need to establish such parameters inspired the present study, especially in light of the asteroid's reputed status. Asteroid 298 Baptistina was therefore observed throughought March and April 2008 from the Abbey Ridge Observatory (Halifax, Canada), the Hunter Hill Observatory (Canberra, Australia), and the Calvin-Rehoboth Observatory (New Mexico, USA).   Details regarding the observatories can be found elsewhere in \citet{la07} \& \citet{ma08} (\textsc{ARO}), \citet{hi06} (\textsc{HHO}), and \citet{mh07} (\textsc{CRO}). Image pre-processing and differential photometry were performed using \textsc{MPO Canopus} \citep{wa06} and \textsc{MaximDL} \citep{ge07}. The asteroid's large proper motion required the selection of different reference stars on each night \citep{wa06,he98}, consequently the \textsc{FALC} algorithm was employed to search both magnitude and temporal space for a period solution \citep{ha89}. The period analysis was carried out in the \textsc{MPO Canopus} \citep{wa06} and \textsc{Peranso} \citep{va07} software environments.

A rotational period of $P_R=16.23\pm0.02$ hours was determined for 298 Baptistina from the analysis, and the resulting phased light curve is presented in Figure 2. The light curve exhibits a peak to peak amplitude of $\simeq0.15$ magnitude and displays complex characteristics that are likely indicative of irregular surface features. Continued photometric observations are envisioned to refine the rotational period, and in conjunction with archival observations by \citet{wi97} and \citet{dh07}, to model the asteroid's shape and spin axis by light curve inversion \citep{mh08,kt01}. The data will also permit a detailed study of the asteroid's absolute magnitude and oppositional surge, fundamental for any subsequent research. Thermal imaging and spectroscopic follow-up would also be of value, permitting a precise determination of the asteroid's diameter and confirmation of its taxonomical class (e.g., \citet{re08}).   This paper's referee has estimated that 298 Baptistina may be \textit{approximately} $\simeq 20$ km in size, which follows from the standard formula utilizing albedo \citep{re08} and the H-magnitude. 

Lastly, the present study appears to reaffirm the importance of small telescopes in conducting pertinent scientific research \citep{pe80,tu05}.  Indeed, modest telescopes can even be mobilized to help address questions surrounding the extinction of the Dinosaurs.
  
\subsection*{ACKNOWLEDGEMENTS}
We are indebted to Petr Pravec, Alan Harris, and Brian Warner for their help in mobilizing the collaboration. DJM also extends his gratitude to the Halifax RASC, Daniel U. Thibault, Aaron Gillich, Joris Van Bever, Arne Henden and the staff at the AAVSO, and Robin Humble and Chris Loken for facilitating simulations of the Kuiper Belt on the McKenzie computer cluster, which is part of the Canadian Institute for Theoretical Astrophysics (CITA) at the University of Toronto.  Lastly, we also thank the referee for his/her comments.


\begin{thebibliography}{}

\bibitem[Bottke et al.(2007)]{bo07} Bottke, W.~F., Vokrouhlick{\'y}, D., \& Nesvorn{\'y}, D.\ 2007, Nature, 449, 48 
\bibitem[Chesley et al.(2003)]{ch03} Chesley, S.~R., et al.\ 2003, Science, 302, 1739 
\bibitem[Ditteon \& Hawkins(2007)]{dh07} Ditteon, R., \& Hawkins, S.\ 2007, Minor Planet Bulletin, 34, 59
\bibitem[George(2007)]{ge07} George D.~B., 2007, \textsc{maximdl} Advanced CCD Imaging Software, http://www.cyanogen.com
\bibitem[Harris et al.(1989)]{ha89} Harris A.~W., et al., 1989, Icarus, 77, 171
\bibitem[Henden \& Kaitchuck(1998)]{he98} Henden A.~A., Kaitchuck R.~H., 1998, Astronomical Photometry: A Text and Handbook for the Advanced Amateur and Professional Astronomer, Willmann-Bell, Richmond 
\bibitem[Higgins et al.(2006)]{hi06} Higgins, D., Pravec, P., Kusnirak, P., Reddy, V., \& Dyvig, R.\ 2006, Minor Planet Bulletin, 33, 64 
\bibitem[Hildebrand(1993)]{hi93} Hildebrand, A.~R.\ 1993, JRASC, 87, 77
\bibitem[Kaasalainen \& Torppa(2001)]{kt01} Kaasalainen, M., \& Torppa, J.\ 2001, Icarus, 153, 24 
\bibitem[Lane(2007)]{la07} Lane D.~J., 2007, 96th Spring Meeting of the AAVSO, http://www.aavso.org/aavso/meetings/spring07present/Lane.ppt
\bibitem[Majaess(2004)]{ma04} Majaess  D.~J.,\ 2004, 
`Assessing the Yarkovsky Effect in the Kuiper Belt', Undergraduate Thesis, Saint Mary's University.
\bibitem[Majaess et al.(2008)]{ma08} Majaess, D.~J., Turner, D.~G., Lane, D.~J., \& Moncrieff, K.~E.\ 2008, Journal of the American Association of Variable Star Observers (JAAVSO), 74 
\bibitem[Molnar \& Haegert(2007)]{mh07} Molnar, L.~A., \& Haegert, M.~J.\ 2007, Minor Planet Bulletin, 34, 126 
\bibitem[Molnar \& Haegert(2008)]{mh08} Molnar, L.~A., \& Haegert, M.~J.\ 2008, AAS/Division of Dynamical Astronomy Meeting, 39, \#02.03 
\bibitem[Percy(1980)]{pe80} Percy, J.~R.\ 1980, JRASC, 74, 334 
\bibitem[Pravec et al.(2002)]{pr02} Pravec, P., Harris, A.~W., \& Michalowski, T.\ 2002, Asteroids III, 113 
\bibitem[Pravec \& Harris(2000)]{ph00} Pravec, P., \& Harris, A.~W.\ 2000, Icarus, 148, 12 
\bibitem[Reddy et al.(2008)]{re08} Reddy, V., et al. 2008, `Composition of 298 Baptistina: Implications for K-T Impactor Link.' Asteroids, Comets, and Meteors Conference, Baltimore Maryland.  
\bibitem[Rubincam(1998)]{ru98} Rubincam, D.~P.\ 1998, Journal of Geophysical Research, 103, 1725 
\bibitem[Turner et al.(2005)]{tu05} Turner, D.~G., Savoy, J., Derrah, J., Abdel-Sabour Abdel-Latif, M., \& Berdnikov, L.~N.\ 2005, PASP, 117, 207 
\bibitem[Vanmunster(2007)]{va07} Vanmunster T., 2007, \textsc{peranso} Light Curve and Period Analysis Software, http://www.peranso.com
\bibitem[Vokrouhlick{\'y} et al.(2006)]{vo06} Vokrouhlick{\'y}, D., Bro{\v z}, M., Bottke, W.~F., Nesvorn{\'y}, D., \& Morbidelli, A.\ 2006, Icarus, 182, 118 
\bibitem[Warner(2006)]{wa06} Warner B.~D., 2006, A Practical Guide to Lightcurve Photometry and Analysis, by B.D.~Warner.~2006 XIII, 297 p.~0-387-29365-5.~Berlin: Springer, 2006  
\bibitem[Wisniewski et al.(1997)]{wi97} Wisniewski, W.~Z., Michalowski, T.~M., Harris, A.~W., \& McMillan, R.~S.\ 1997, Icarus, 126, 395 
\end{thebibliography}
\end{document}